\documentclass[%
 reprint, 
 amsmath,amssymb,
 aps, physrev,
floatfix,
]{revtex4-2}

\usepackage{url}
\usepackage[T1]{fontenc}
\usepackage[caption=false]{subfig}
\usepackage{graphicx}
\usepackage{dcolumn}
\usepackage{bm}
\usepackage[colorlinks=true, linkcolor=blue, citecolor=blue, urlcolor=blue]{hyperref} 

\usepackage{tikz}
\usetikzlibrary{arrows.meta}
\usetikzlibrary{decorations.pathreplacing}
\usepackage{placeins}

\begin{document}

\title{Chiral-split magnons in the S = 1 Shastry-Sutherland model}
\author{Absur Khan Siam}
\author{Se Kwon Kim}
\affiliation{Department of Physics, Korea Advanced Institute of Science and Technology, Daejeon 34141, Republic of Korea}
\date{\today}

\begin{abstract}
     In ferromagnets, magnons have only one chirality; while in common antiferromagnets, bands with opposite chiralities are degenerate across the Brillouin zone. Recent studies have shown that it is possible to observe non-degenerate bands of opposite chiralities in altermagnetic materials. Here we take the S = 1 Shastry-Sutherland model, which shows the collinear N\'eel (I) phase, and investigate the magnon band structure showing alternate chirality-splitting and the resulting transport properties. In magnon bands, we find a notable feature of the chirality-split magnon bands, and the split is opposite along two different directions in the Brillouin zone. We also calculate the spin and thermal conductivities using Kubo formalism. Our calculations show robust spin Seebeck and spin Nernst effects due to the alternating chirality split across the Brillouin zone, without any external magnetic field and spin-orbit coupling.
\end{abstract}

\maketitle

\section{Introduction}

The interplay between symmetry and quasiparticle dynamics plays a pivotal role in shaping the transport properties of magnetic materials. In ferromagnets, magnonic excitations inherently exhibit a single right-handed chirality due to the broken time-reversal symmetry~\cite{liu2022switching, PhysRevLett.125.027201, PhysRevLett.133.166705}. In contrast, conventional collinear antiferromagnets, especially those with easy-axis anisotropy, host degenerate magnon modes carrying opposite spin angular momenta, protected by combined symmetry operations such as time-reversal and spatial inversion~\cite{brinkman1966theory, rezendeMagnonSpintronicsReview, PhysRevB.89.081105, li2020spin, PhysRevLett.118.147202, PhysRevLett.104.217204, PhysRevLett.124.217201}. Despite having vanishing net magnetization and static stray fields, such antiferromagnets have been actively explored for spin-based nanoscale devices, including spin Seebeck and spin Nernst effects, which transport spins longitudinally and transversely in response to thermal gradients~\cite{PhysRevLett.118.147202, PhysRevLett.124.217201}.

However, these effects often require either an applied magnetic field to lift spin degeneracy~\cite{PhysRevLett.104.217204} or intrinsic spin-orbit coupling (SOC) to induce Berry curvature~\cite{PhysRevLett.124.217201}. Theoretical proposals have tried to bypass these constraints by exploring noncollinear antiferromagnets such as kagome lattices~\cite{PhysRevB.105.174401}, but these systems may still rely on Dzyaloshinskii-Moriya interactions (DMI) or field-induced spin configurations to enable efficient spin transport~\cite{PhysRevB.111.064401}. Thus, realizing robust magnonic spin transport in antiferromagnets, completely free of external fields and relativistic SOC, remains a central challenge.

Recent advances have discovered a novel class of materials termed \textit{altermagnets}, which defy the traditional dichotomy between ferromagnets and antiferromagnets~\cite{PhysRevX.12.031042}. These materials, despite exhibiting no net magnetization, can host nondegenerate magnon modes with opposite chiralities, enabled purely by symmetry considerations: specifically, the breaking of combined inversion, time-reversal, and translational symmetries~\cite{PhysRevLett.131.256703, PhysRevLett.133.156702, PhysRevB.107.104408}. In contrast to mechanisms relying on SOC or DMI, altermagnetic systems offer an intrinsically symmetry-protected route to chiral magnon transport and thermally driven spin currents. While metallic altermagnets such as RuO\textsubscript{2} have demonstrated spin-split bands and spin Hall-like behavior~\cite{PhysRevX.12.031042}, insulating altermagnets offer a cleaner setting to explore magnon-mediated spin transport without complications from charge carriers.

Among the proposed platforms for altermagnetic behavior, the Shastry-Sutherland lattice has recently gained attention. In the spin-1/2 case, the N\'eel phase of the Shastry-Sutherland lattice was shown to exhibit altermagnetic features~\cite{PhysRevB.110.205140}. However, the spin-1 version of the model, known to stabilize a collinear N\'eel (I) phase~\cite{doi:10.1143/JPSJ.72.938}, still remains unexplored in this context.

In this work, we demonstrate that the collinear N\'eel phase of the $S = 1$ Shastry-Sutherland model possesses altermagnetic character. We establish this both from symmetry considerations and from a direct analysis of magnon dispersions. Our study reveals a unique form of chirality-splitting in the magnon bands: the splitting reverses sign along two different directions in momentum space. Furthermore, we compute the resulting spin Seebeck and spin Nernst coefficients using the Kubo formalism and show that these effects remain robust without the need for spin-orbit coupling or applied magnetic fields. Our findings enrich the growing landscape of altermagnetic phenomena and suggest new pathways for controlling magnonic transport in antiferromagnetic systems.

\section{Model and Methods}
We consider the generalized Shastry-Sutherland model. The spin Hamiltonian is given by \cite{doi:10.1143/JPSJ.72.938}
\begin{equation}
    H = \sum_{(i,j)}J_{ij}\mathbf{S}_i\cdot \mathbf{S}_j + D\sum_i (S_i^z)^2,
    \label{eq:SSL_model}
\end{equation}
where $\mathbf{S}_i$ denotes the spin operator on the $i$th site. As illustrated in Fig.~\ref{subfig:SS_a} $J_{ij} = J$, $J'$, and $J''$ represent the intra-dimer, the inter-dimer and the inter-chain couplings, which are all assumed to be antiferromagnetic. The single-ion anisotropy is denoted as $D <0$.

\begin{figure}

    \input{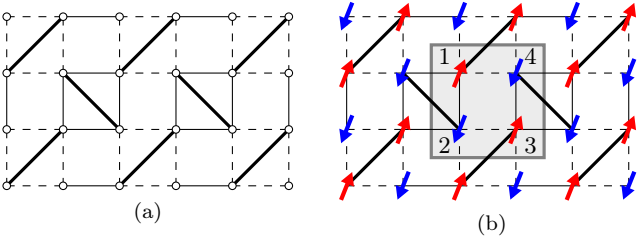}

    \caption{Generalized Shastry-Sutherland model. (a) Orthogonal-dimer structure: the bold, thin and broken lines represent the exchange couplings $J$, $J'$ and $J''$. Note that the dimers indicated by the bold lines are orthogonal to each other. (b) Spin configurations for the N\'eel ordered phase (I). Unit cell is denoted by the gray square: sites 1 and 3 have spin up and sites 2 and 4 have spin down.}

\end{figure}

The N\'eel (I) phase (Fig.~\ref{subfig:SS_b}) of the orthogonal-dimer structure belongs to the plane group $p4g$: the 4-fold axes are at the center and corner of the unit cell, and the 2-fold axes (inversion centers) are at the center of the edges of the cell. We verify the N\'eel (I) phase to be altermagnetic by checking the identification rules \cite{PhysRevX.12.040501}:

\begin{enumerate}
    \item There is an even number (four) of magnetic atoms in the unit cell, and the number of atoms in the unit cell does not change between the nonmagnetic and magnetic phases of the crystal.
    \item There is no inversion center between the sites occupied by the magnetic atoms from the opposite-spin sublattices. Any inversion with respect to the inversion centers maps to the same sublattice.
    \item The two opposite-spin sublattices are connected by crystallographic rotation transformation $C_{4z}$.
\end{enumerate}

The existence of a N\'eel (I) phase as the ground state of~\eqref{eq:SSL_model} has been shown for a range of parameters~\cite{doi:10.1143/JPSJ.72.938}.

To address quantum fluctuations around the ground state, we reformulate the Hamiltonian in terms of the Holstein-Primakoff (HP) transformation \cite{PhysRev.58.1098}, utilizing the creation (annihilation) operators $b_i^\dag$ ($b_i$). 
The unit cell consists of four sites; in our labeling, sites 1 and 3 have spin up and sites 2 and 4 have spin down in the ground state. Thus, for sites 1 and 3, we have $S^+ = (2S - b^\dag b)^{1/2}b,~S^z = S - b^\dag b$, and for sites 2 and 4, $S^+ = b^\dag (2S - b^\dag b)^{1/2},~ S^z = -S + b^\dag b$.
Within the framework of the linear spin wave theory, we only keep the quadratic terms. The resulting Hamiltonian is then transformed into the reciprocal space using a Fourier transformation, $b_i^\dagger = \frac{1}{\sqrt{N}}\sum_{\mathbf{k}}^{\rm B.Z.}e^{-i\mathbf{k}\cdot\mathbf{R}_i}b_{\mathbf{k}}^\dagger$, where $N$ is the number of unit cells. The Hamiltonian in the reciprocal space is given by 
\begin{equation}
    H = \frac{1}{2}\sum_\mathbf{k}\Psi_\mathbf{k}^\dag\mathcal{H}_\mathbf{k}\Psi_\mathbf{k},
\end{equation}
where the Nambu spinor $\Psi_\mathbf{k} = \left({b^1_\mathbf{k}}, ~{b^2_\mathbf{k}},~{b^3_\mathbf{k}}, ~{b^4_\mathbf{k}}; ~{b_{-\mathbf{k}}^1}^\dag, ~{b_{-\mathbf{k}}^2}^\dag, ~{b_{-\mathbf{k}}^3}^\dag, ~{b_{-\mathbf{k}}^4}^\dag\right)^T$, with ${b^l_\mathbf{k}}^\dag$ $(l = 1,~2,~3,~4)$ being the creation operator defined at the four sites in the unit cell. The Hamiltonian matrix in the reciprocal space takes the form of
\begin{equation}
    \mathcal{H}_\mathbf{k} = S\begin{pmatrix}
    M(\mathbf{k}) & N(\mathbf{k})\\
    N(\mathbf{k}) & M(\mathbf{k})
    \end{pmatrix},
\end{equation}
where $M(\mathbf{k})$ and $N(\mathbf{k})$ are $4\times4$ Hermitian matrices:
\begin{align*}
    M(\mathbf{k}) &= \begin{pmatrix}
    \Xi & 0 & Je^{-ik_ya} & 0\\
    0 & \Xi & 0 & Je^{ik_xa}\\
    Je^{ik_ya} & 0 & \Xi & 0\\
    0 & Je^{-ik_xa} & 0 & \Xi\\
    \end{pmatrix},\\
    N(\mathbf{k}) &= \begin{pmatrix}
    0 & \alpha & 0 & \beta\\
    \alpha^* & 0 & \beta & 0\\
    0 & \beta^* & 0 & \alpha^*\\
    \beta^* & 0 & \alpha & 0\\
    \end{pmatrix},
\end{align*}
with $\alpha = J'' + J'e^{ik_xa}$, $\beta = J' + J''e^{ik_ya}$, and $\Xi = 2J' + 2J'' - J - 2D$; $a$ is the lattice constant.

The eigenpairs of the magnonic Hamiltonian are obtained by the Bogoliubov transformation $\psi_\mathbf{k} = T_\mathbf{k}\Psi_\mathbf{k}$ which conserves the commutation relation of bosonic operators. We diagonalize the Hamiltonian matrix using the Cholesky decomposition \cite{COLPA1978327,PhysRevB.87.174427} and obtain the eigenenergies.

\begin{figure}
    \centering
    \includegraphics[width = 0.8\columnwidth]{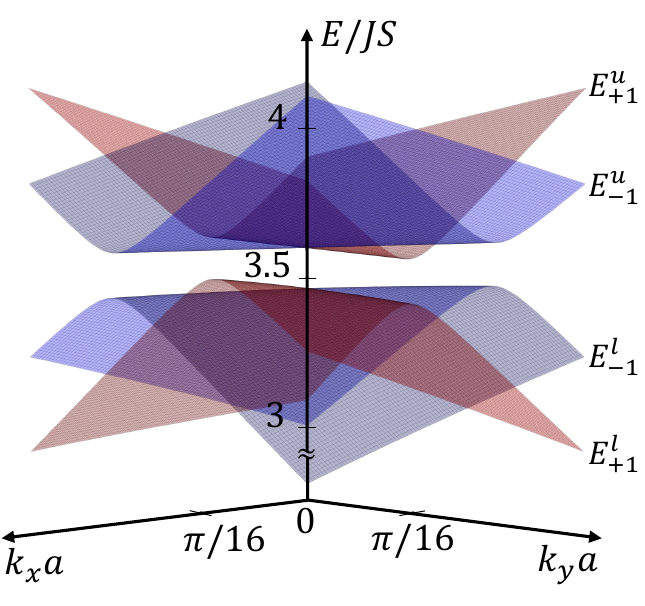}
    \caption{Magnon band dispersion of the generalized Shastry-Sutherland model near $\mathbf{k} = 0$. Red and blue indicate the left and right-handed magnon modes, respectively. The values $J'/J = 1.2$, $J''/J = 1$, and $D/J = -0.6$ are used for the parameters.}
    \label{fig:magnon_3d}
\end{figure}

A distinct chirality, corresponding to the direction of precession of the spin around the axis of the collinear magnetic order, can be assigned to each of the magnon bands. Since the collinear order in the $z$-direction preserves the $z$-component of the total spin $S_z = \sum_i S_i^z$ with $i$ running over all sites, we use $\vartheta = \langle S^z\rangle/\hbar$ to specify the chirality of the bands \cite{PhysRevLett.117.217202}. In particular, we have $\vartheta = -1$ for the right-handed and $\vartheta = +1$ for left-handed magnon modes. In other words, magnon modes with chirality $\vartheta = \pm 1$ carry $\pm 1$ spin angular momentum along the $z$-direction. Since there are distinct upper and lower bands in the dispersion with opposite chiralities in each, we label the bands as $E^u_{\pm 1}(\mathbf{k})$ and $E^l_{\pm 1}(\mathbf{k})$, respectively, where the subscript denotes the chirality $\vartheta$ of the band (Fig.~\ref{fig:magnon_3d}).

The magnon spin current can be driven by a temperature gradient. In the linear response regime, the magnon spin current and temperature gradient are related by
\begin{equation}
    \begin{pmatrix}
            j_x^z \\ j_y^z
    \end{pmatrix} = 
    \begin{pmatrix}
        \sigma_{xx} & \sigma_{xy}\\
        \sigma_{yx} & \sigma_{yy}
    \end{pmatrix}
    \begin{pmatrix}
        (-\partial_xT)\cos{\theta} \\ (-\partial_yT)\sin{\theta}
    \end{pmatrix},
\end{equation}
where $\theta$ is the angle between the temperature gradient $\nabla T$ and the $\hat{\mathbf{x}}$ direction. Based on the Kubo formula~\cite{PhysRevB.108.L180401, PhysRevB.100.100401, kubo1957statistical, Coleman_2015, PhysRevB.89.054420, PhysRevB.99.174402}, we can calculate the magnon conductivity tensor $\sigma_{mn} = \sum_\mu\sigma_{mn}^\mu$ where $\sigma_{mn}^\mu$ is the contribution from the mode $\mu \in \{E^u_{\pm 1},~E^l_{\pm 1}\}$
\begin{equation*}
    \sigma_{mn}^\mu = (\hat{S}^z)_{\mu\mu}\frac{\tau_0}{A \hbar k_B T^2} \sum_\mathbf{k} ( \hat{V}_m )_{\mu\mu} ( \hat{V}_n )_{\mu\mu} \frac{\mu_\mathbf{k} e^{-\mu_\mathbf{k} / k_B T}}{( e^{\mu_\mathbf{k} / k_B T} - 1 )^2},
\end{equation*}
where $n$ is the direction of the temperature gradient, $m$ is the direction of the spin current, $A$ is the sample area, $\tau_0$ is the magnon lifetime, $\hat{V}_m = T_\mathbf{k}^\dagger (\partial_{k_m} \mathcal{H}_\mathbf{k}) T_\mathbf{k}$, and $\mu_\mathbf{k} = \mu(\mathbf{k})$ is the dispersion relation of the mode $\mu$.

We define a rotated coordinate system such that the thermal gradient is aligned with the new $x'$-axis, at an angle $\theta$ with respect to the original $x$-axis. The rotation matrix is
\begin{equation*}
    R(\theta) = 
    \begin{pmatrix}
        \cos\theta & \sin\theta \\
        -\sin\theta & \cos\theta
    \end{pmatrix}.
\end{equation*}

Under this rotation, the thermal conductivity tensor transforms as
\begin{equation}
    \bm{\sigma}' = R(\theta) \, \bm{\sigma} \, R(\theta)^{-1},
\end{equation}
where $\bm{\sigma}$ is the original thermal conductivity tensor, and $\bm{\sigma}'$ gives the components in the rotated frame. In this rotated frame, the diagonal component $\sigma'_{x'x'}$ corresponds to the spin Seebeck response (along the gradient)~\cite{adachi2013theory}, and the off-diagonal component $\sigma'_{y'x'}$ corresponds to the spin Nernst response (transverse to the gradient)~\cite{PhysRevB.84.184406, PhysRevLett.97.026603, go2024magnon}.

The thermal gradient is applied at an angle $\theta$ to the $x$-axis, i.e., $\nabla T = |\nabla T|(\cos\theta, \sin\theta)^T,$ then the spin current projected \emph{along} and \emph{perpendicular} to the thermal gradient is given by:
\begin{equation}
    \begin{pmatrix}
        j_\parallel^z \\
        j_\perp^z
    \end{pmatrix}
    =
    -|\nabla T|
    \begin{pmatrix}
        \sigma_\parallel \\
        \sigma_\perp
    \end{pmatrix},
\end{equation}
where
\begin{align*}
    \sigma_\parallel &= 
        \Phi\cos\theta +
        \Omega\sin\theta, \\
    \sigma_\perp &= 
        -\Phi\sin\theta +
        \Omega\cos\theta,
\end{align*}
where $\Phi = \sigma_{xx}\cos\theta + \sigma_{xy}\sin\theta$ and $\Omega = \sigma_{yx}\cos\theta + \sigma_{yy}\sin\theta$. Here, $\sigma_\parallel$ is the \textit{spin Seebeck coefficient} (longitudinal response), and $\sigma_\perp$ is the \textit{spin Nernst coefficient} (transverse response).

\section{Results}
\begin{figure}
    \subfloat[\label{fig:split1}]{\includegraphics[width=.5\columnwidth]{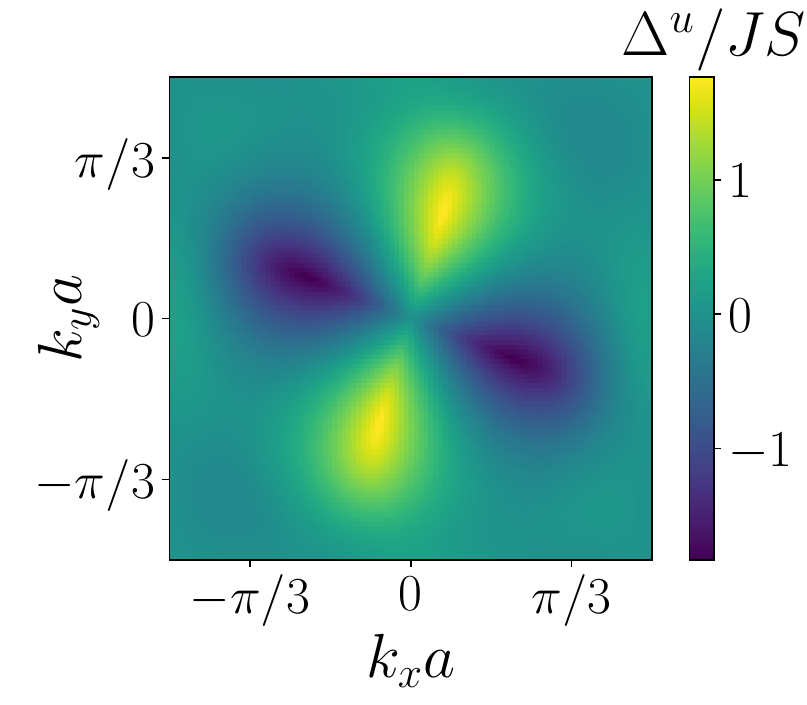}}
    \subfloat[\label{fig:split2}]{\includegraphics[width=.5\columnwidth]{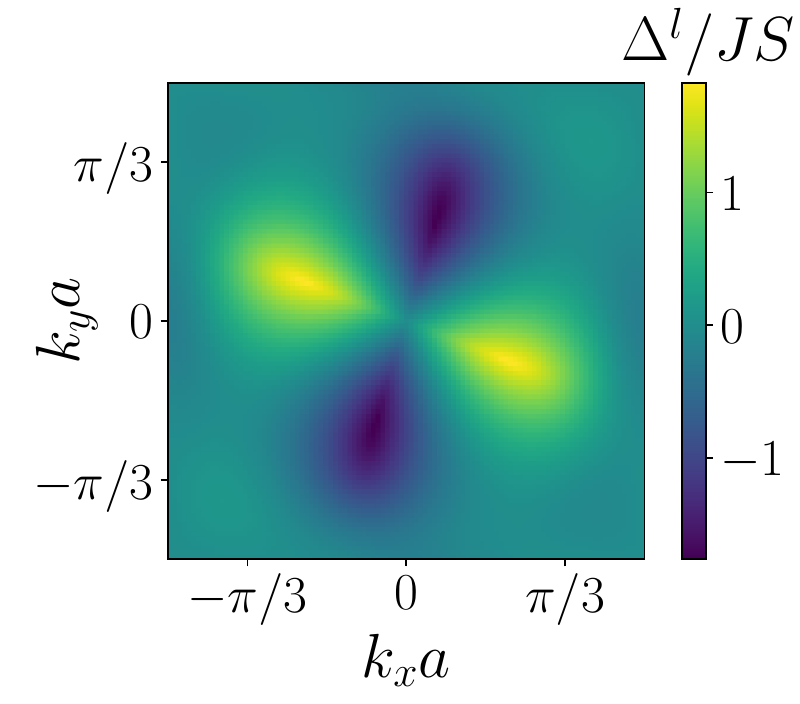}}
    \qquad
    \subfloat[\label{fig:dispersion}]{\includegraphics[width=0.5\columnwidth]{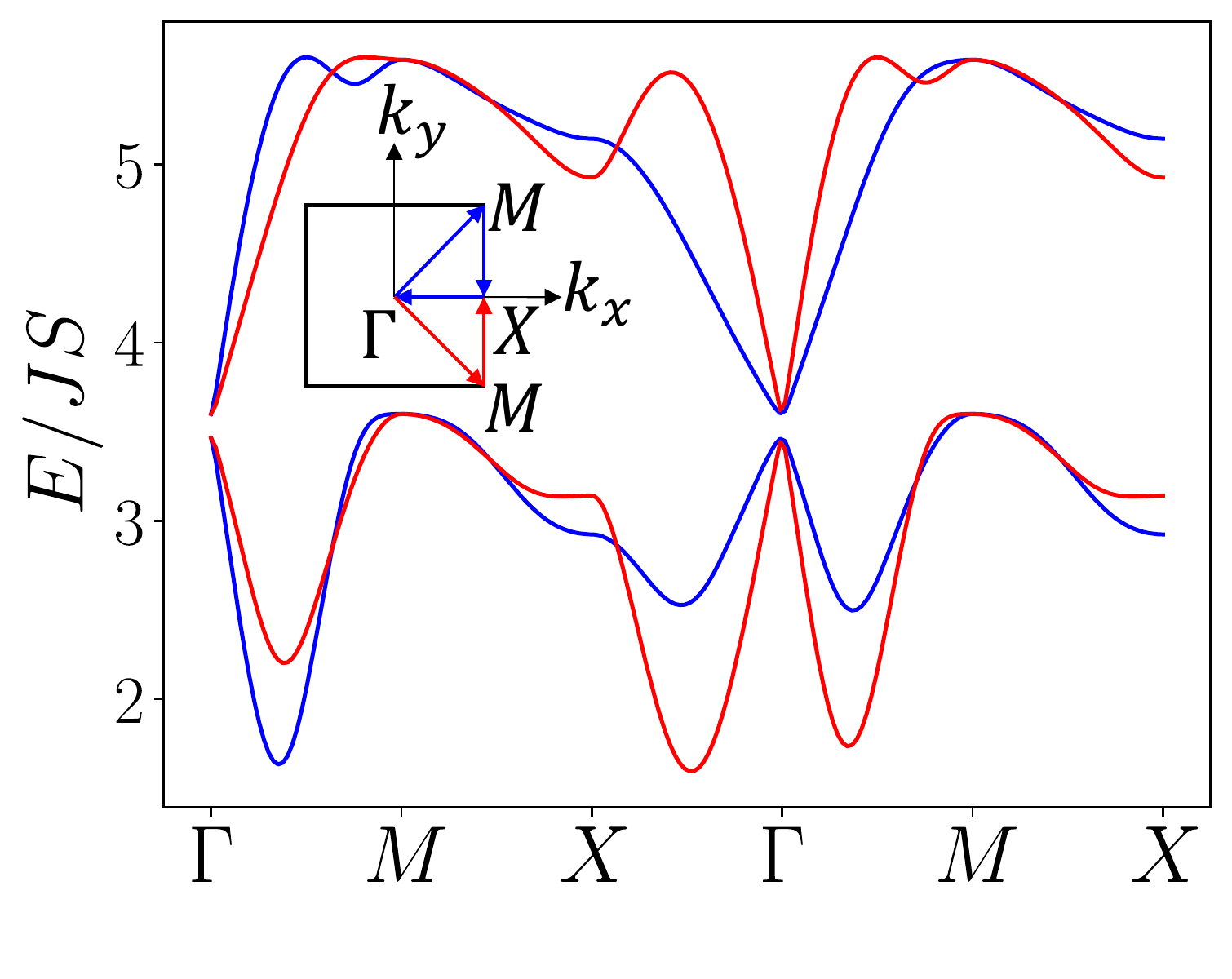}}
    \subfloat[\label{fig:spin_seebeck_nernst}]{\includegraphics[width=0.5\columnwidth]{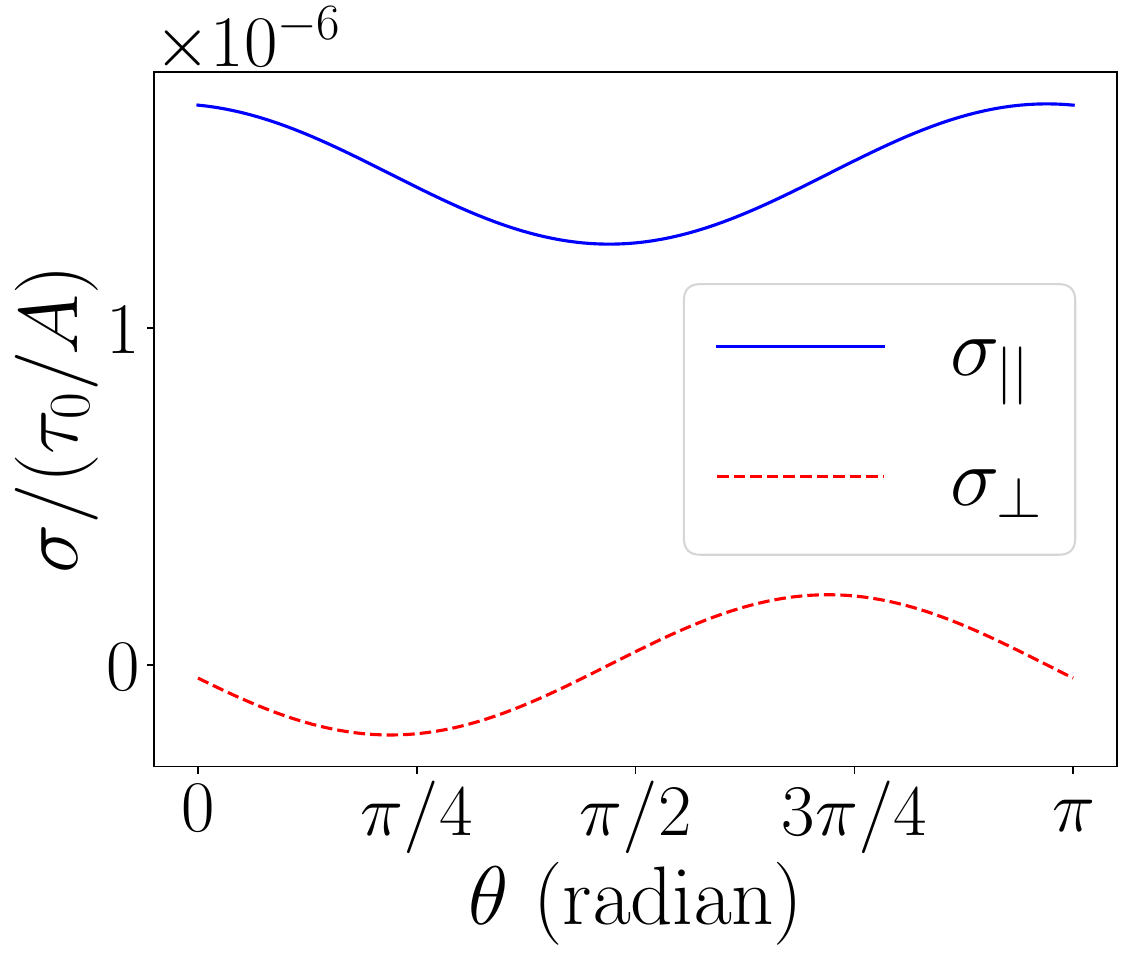}}
    \caption{(a) Energy difference between upper bands with right and left-handed chirality, (b) energy difference between lower bands with right and left-handed chirality, (c) chirality-split dispersion relation of magnon along the path $\Gamma \to M \to X \to \Gamma \to M \to X$ shown in the inset. We first trace the path $\Gamma \to M \to X \to \Gamma$ (blue arrows), and then $\Gamma \to M \to X$ (red arrows). Red and blue denote the left and right-handed magnon modes, respectively. Here we used $J'/J = 1.2$, $J''/J = 1$, and $D/J = -0.6$. (d) Variation of spin Seebeck ($\sigma_{\parallel}$) and spin Nernst ($\sigma_{\perp}$) coefficients for different values of $\theta$. Here, in addition, we used $J/k_B = 100\text{ K}$ and $T = 25\text{ K}$.}
\end{figure}

The magnon dispersion is shown in Fig.~\ref{fig:magnon_3d}. We have used the parameter values $J'/J = 1.2$, $J''/J = 1$, and $D/J = -0.6$ unless otherwise stated. There are four bands: two with opposite chiralities with higher energy, $E^u_{\pm 1}$, and two with opposite chiralities with lower energy, $E^l_{\pm 1}$, with a strict energy gap between these two sectors. We will refer to these two pairs of bands as the lower and the upper bands, respectively.

We observe a $C_2$ symmetry in all of the magnon bands, as expected. To observe the chirality split, we calculated the energy difference between the upper bands $\Delta^u(\mathbf{k}) = E^u_{- 1} - E^u_{+ 1}$ and between the lower bands $\Delta^l(\mathbf{k}) = E^l_{- 1} - E^l_{+ 1}$ in Fig.~\ref{fig:split1} and Fig.~\ref{fig:split2}, respectively. We also plot the dispersion along a path with high-symmetry points (Fig. \ref{fig:dispersion}). From these we can confirm that we have opposite chirality split along two different directions.


We then calculated the spin Seebeck ($\sigma_{\parallel}$) and spin Nernst ($\sigma_{\perp}$) coefficients with $J/k_B = 100\text{ K}$ and $T = 25\text{ K}$. The angular variation of the spin Seebeck and Nernst coefficients arises from the anisotropic nature of the spin conductivity tensor. In Fig.~\ref{fig:spin_seebeck_nernst}, the spin Seebeck coefficient $\sigma_{\parallel}$ exhibits a broad minimum around $\theta = \pi/2$, where the thermal gradient is directed along the $y$-axis. This minimum results from the fact that $\sigma_{yy} < \sigma_{xx}$, suppressing longitudinal transport in the $y$ direction. Conversely, $\sigma_{\parallel}$ peaks near $\theta = 0$ and $\pi$, where the thermal gradient aligns with the $x$-axis and the contribution from $\sigma_{xx}$ is dominant. 

The spin Nernst coefficient $\sigma_{\perp}$ exhibits a sinusoidal angular dependence and vanishes at $\theta = 0,~\pi/2,~\pi$, where the thermal gradient is aligned with the crystal axes. At these angles, the transverse projection of the spin current cancels due to geometric symmetry. The extrema near $\theta = \pi/4$ and $3\pi/4$ arise when the thermal gradient is oriented diagonally, allowing both $\sigma_{xx}$ and $\sigma_{yy}$ to contribute significantly to the transverse response. Since the off-diagonal term $\sigma_{xy}$ is more than an order of magnitude smaller than the diagonal terms, the shape of $\sigma_{\perp}(\theta)$ is primarily governed by the anisotropy between $\sigma_{xx}$ and $\sigma_{yy}$, with $\sigma_{xy}$ introducing only a small asymmetry between the maximum and minimum.

We have checked that the spin Nernst coefficient is uniformly zero for the case $J' = J''$ as expected, since the $x$ and $y$ directions are related by a $90$ degree rotation in this case.

\section{Conclusion}
In this work, we have demonstrated that the $S=1$ Shastry-Sutherland model in the collinear N\'eel (I) phase exhibits altermagnetic behavior, characterized by magnon bands of opposite chiralities that are non-degenerate across the Brillouin zone. Through linear spin-wave analysis, we identified a unique pattern of chirality splitting, distinctively opposite along different directions, arising purely from the symmetry of the model, without requiring spin-orbit coupling or external magnetic fields.

The chirality-resolved magnon bands show a robust $C_2$ symmetry, and the energy differences between bands of opposite chirality reveal clear signatures of alternating splitting. This chirality-split band structure results in finite spin Seebeck and spin Nernst responses, calculated using the Kubo formalism. Our results demonstrate that the spin Seebeck effect persists for all orientations of the thermal gradient, while the spin Nernst effect vanishes along high-symmetry directions.

These findings suggest that altermagnetic antiferromagnets such as the $S=1$ Shastry-Sutherland system can host unconventional magnon transport phenomena and provide a promising platform for spin caloritronic applications without requiring relativistic effects. Future work may explore the tunability of these transport signatures by the variation of the exchange couplings.

In particular, rare-earth melilite compounds such as $\mathrm{Pr_2Be_2GeO_7}$ and $\mathrm{Pr_2Ga_2BeO_7}$, which realize the Shastry-Sutherland geometry, offer promising material platforms for experimental investigation~\cite{doi:10.1021/acs.inorgchem.0c03131}. The Pr$^{3+}$ ions in these systems can support effective spin-1 moments depending on the local crystal field environment, making them relevant candidates for realizing altermagnetic behavior and exploring spin-dependent thermal transport. We note that our semiclassical linear spin-wave treatment does not rely on the specific value of spin $S$, and remains valid for higher-spin systems as long as the system remains in the N\'eel phase.

\section{Acknowledgment}
 A.K.S. thanks Daehyeon An and Seungho Lee for many helpful discussions. This work was supported by Brain Pool Plus Program through the National Research Foundation of Korea funded by the Ministry of Science and ICT (2020H1D3A2A03099291), National Research Foundation of Korea (NRF) grant funded by the Korea government(MSIT) (2021R1C1C1006273), and Basic Science Research Program through the National Research Foundation of Korea (NRF) funded by the Ministry of Education (2019R1A6A1A10073887).

\bibliographystyle{apsrev4-2}
\bibliography{ref}

\end{document}